# Understanding Photo-thermal and Melting Mechanisms in Optical Charging of Nano and Micro Particles Laden Organic PCMs

Domala Sai Suhas [a] and Vikrant Khullar[a,*]
[a]Mechanical Engineering Department, Thapar Institute of Engineering & Technology, Patiala-147004, Punjab, India

*Corresponding author. Email address: vikrant.khullar@thapar.edu

**ABSTRACT**
The realm of latent heat storage has witnessed emergence of optical charging as a promising route of solar thermal latent heat storage. However, it is still in its initial stages of development and warrants further investigations to take it to the next level i.e., realization of optical charging based real-world systems. Engineering efficient optical charging process in turn necessitates efficient photo-thermal energy conversion, transfer as well as storage of the incident solar radiant energy. The present work is a determining step in deciphering, quantifying, and understanding the aforementioned steps involved in the optical charging process. In particular, experiments have been designed carefully to investigate optical charging of composite-PCMs (particles laden organic PCMs) with and without thermochromism assistance. Spatial-temporal temperature distribution curves reveal that temperature spread (in the liquid phase) in case of optical charging of non-thermochromic particles (carbon soot nanoparticles) laden PCMs is significantly high (as high as approximately 24 °C) relative to that observed in case of thermochromic particles (microcapsules) laden PCMs (approximately, 4 °C). The magnitude of the temperature spread (being representative of the deviation from thermostatic optical charging) clearly points out that opposed to non-thermochromic laden PCMs, nearly thermostatic optical charging can be achieved in case of thermochromic particles laden PCMs. Furthermore, in case of optical charging without thermochromic assistance, the temperature spread, peak temperatures and the melting rates increase with increase in particles concentration. Whereas, in the latter case, although the temperature spread and peak temperatures are nearly independent; the melting rates do depend on the particles concentration.

## 1. INTRODUCTION

Latent heat storage is a straightforward, efficient and inexpensive route of storing solar thermal energy. However, the inherently low thermal conductivity of phase change materials (PCMs) and low convective heat transfer coefficient values limit the charging and discharging rates. Various approaches have been devised to improve these aspects; for instance, numerous researchers have explored adding high thermal conductivity materials (particularly nanoparticles) into the PCM [1-4] and or employing carbon/metal meshes/sheets to improve the thermal conductivity [5-9]. Furthermore, to improve the convective heat transfer rates, researchers have focused on employing fins, devising mechanisms that result in efficient mixing and motion of the liquid phase [10-12].
Although the aforementioned approaches do improve the effective thermal conductivity and heat transfer coefficient values, the enhancements in terms of the melting rates are not significantly high. Recently, an alternative approach, popularly known as 'optical charging' (involving radiative heat transfer as the primary mechanism) has emerged as a promising route of latent heat storage. Herein, the PCM is seeded with broad absorption-based nanoparticles (PCM seeded with nanoparticles is often called nano-PCM) and allowed to directly and volumetrically interact with the incident sunlight - considerably reducing the intermediate thermal resistances and thus resulting in accelerated melting of the PCM. However, this accelerated melting gradually slows down owing to the fact that as the melting progresses, the incident radiations have to transverse through the optically opaque liquid (melted nano-PCM) in order to reach the solid nano-PCM layer [13-16]. Recently, few researchers have tried to address this issue through seeding magnetic nanoparticles (instead of non-magnetic nanoparticles) into the PCM and employing external magnetic field. In particular, magnetic field was employed to displace the nanoparticles in the liquid phase to make the way for incident radiations to interact with the solid nano-PCM front [13, 16]. Although effective, the aforementioned approach is energy intensive and practically difficult to realize and scaleup. As an alternative approach, very recently, the concept of thermochromism-assisted optical charging has been explored. Herein, an otherwise optically transparent (both in solid and liquid phases) PCM seeded with thermochromic nanoparticles was analysed. Theoretical modelling results reveal that the melting rate of PCM could be significantly accelerated (approx. 168 % faster relative to the conventional thermal charging route) through thermochromism assistance at nearly thermostatic conditions [17]. However, there are many instances when the PCM is optically opaque in the solid phase but optically transparent in the liquid phase; for instance, most organic PCMs have the aforementioned optical characteristics. In this direction, the only reported study is that of Ren et. al., [18] wherein thermochromic PCM was synthesized through addition of thermochromic compound (2-anilino-6-dibutylamino-3-methylfluoran (ODB-2) and bisphenol A (BPA)) into an organic PCM (1-hexadecanol (1-HD)). Herein, the synthesis, characterization, and proof of the concept experiments pertinent to photothermal behaviour of the synthesized thermochromic PCM were reported. However, no data or analysis with regard to the spatial temperature distribution and melting front was reported. On the whole, the concepts of optical charging with and without assistance have been recently explored by various researchers (see Fig. 1). However, more holistic and systematic studies are warranted to better understand the photo-thermal behaviour and melting mechanisms, so that real world dispatchable systems (based on the aforementioned concepts) could be realized.

The present work is a significant step in this direction. In particular, comprehensive experimental investigations have been carried out to understand the effect of adding thermochromic particles on the melting front evolution and spatial (as well as temporal) temperature distribution within the thermochromic PCM. Furthermore, to clearly delineate and understand the differences (in terms of photo-thermal behaviour and melting mechanisms) between optical charging routes with and without thermochromism assistance; the case of optical charging of non-thermochromic PCM has also been critically analysed. Overall, the present work is a determining step towards realizing nearly thermostatic and accelerated optical charging of PCMs.

| Pristine PCM (optical characteristics) | OPTICAL CHARGING | | |
|---|---|---|---|
| | No dynamic control of additive particles' optical properties | Dynamic control of additive particles' optical properties | |
| | | Magnetic field assistance | Thermochromism assistance |
| Solid phase: Optically transparent / Liquid phase: Optically transparent | Wang et. al., 2017 [E, T] | Wang et. al., 2017 [E, T] | Singh and Khullar, 2023 [T] |
| Solid phase: Optically opaque / Liquid phase: Optically transparent | Du et. al., 2019 [E] Moudgil and Khullar 2019 [E] Shi et. al., 2020 [E] | Shi et. al., 2020 [E] | Ren et. al., 2019 [E] |

[E]: Experimental modelling, [T]: Theoretical modelling    SCOPE OF THE PRESENT WORK

Fig. 1 List of selected studies pertinent to optical charging of organic PCMs and the scope of the present work.

## 2. IDENTIFYING THE KEY PARAMETERS IMPACTING THE OPTICAL CHARGING PROCESS

In optical charging process, the incident solar irradiation directly interacts with the PCM. Therefore, the melting rate and hence the latent heat storage capacity are highly dependent on the optical properties of the PCM.

As noted earlier, in their pristine form, the organic PCMs are optically opaque (particularly scatter in the visible region) in the solid phase and are optically transparent in the liquid phase. Therefore, to improve their absorption capability (and hence the melting rate), it is imperative to seed them with highly absorbing particles – hereon the aforementioned additive particles laden PCM shall be referred to as "composite-PCM". Furthermore, once a portion of the aforementioned composite PCM has melted, the incident radiations have to pass through the melted layer to reach the solid interface. Therefore, in the initial stages of melting, the optical characteristics of the solid phase of "composite PCM" shall play the predominant role, whereas as the melting progresses, the optical characteristics of liquid phase of "composite PCM" shall start to play an increasing important role in dictating the melting process. Ideally, once a layer of composite PCM has melted, it should then allow the radiation to pass through it and strike the next solid layer. This shall result in nearly thermostatic and accelerated optical charging of the composite-PCM (see Fig. 2(b)).

In actual practice, the solar radiations interact with these particles through absorption and scattering mechanisms (see Fig. 2(b)). Therefore, the relative magnitudes of the aforementioned mechanisms shall dictate the extent to which the ideal thermostatic optical charging could be achieved.

The relative magnitudes of absorption and scattering in turn depends on particles' material, size, shape, and volume fraction. In relation to particles' material, any material which doesn't chemically react with the PCM and has high absorptivity in the incident solar wavelength band qualifies to be a good additive material. In particular, carbon-based materials (in various allotropic forms) have been extensively researched owing to their broad absorption characteristics, enhanced thermo-physical properties, and at the same time being inexpensive. More recently, thermochromic materials (particularly leuco dye based) have been shown to possess temperature dependent optical properties. Tailoring transition temperature ($T_{tr}$) of the thermochromic particles close to the PCM's melting temperature shall assist in making way for the incident radiation through the molten layer to reach the next solid interface. Furthermore, for a given material, it is essentially the particle morphology (size and shape) that dictates the relative magnitudes of scattering and absorption of the incident radiation. Particle size may vary from few nanometres to a few micrometres, the particle shape may be standard (spherical, cylinder, etc.) or amorphous. Figure 2(c) shows the optical characteristics displayed by various additive particles,

Ideally, in the solid phase, the composite-PCM should absorb all the incident solar radiations. On the other hand, in the liquid phase it should transmit all the solar radiations so that the incident radiations are able to strike the fresh layer of solid composite-PCM. Clearly, for a given material and shape, nanoparticles have higher absorption capability and scatter less relative to their micron sized counterparts. Furthermore, for a given size and shape, thermochromic particles promise greater penetration depths relative to their non-thermochromic counterparts. With the purpose of understanding the effect of thermochromism, absorption, and scattering mechanisms on the photo-thermal energy conversion and melting process on the whole, two cases have been investigated. Firstly, the best-case scenario in the realm of non-thermochromic optical charging (wherein non-thermochromic nanoparticles laden PCM is optically heated), secondly, the worst-case scenario in the realm of thermochromic optical charging (wherein thermochromic microcapsules laden PCM is optically heated).

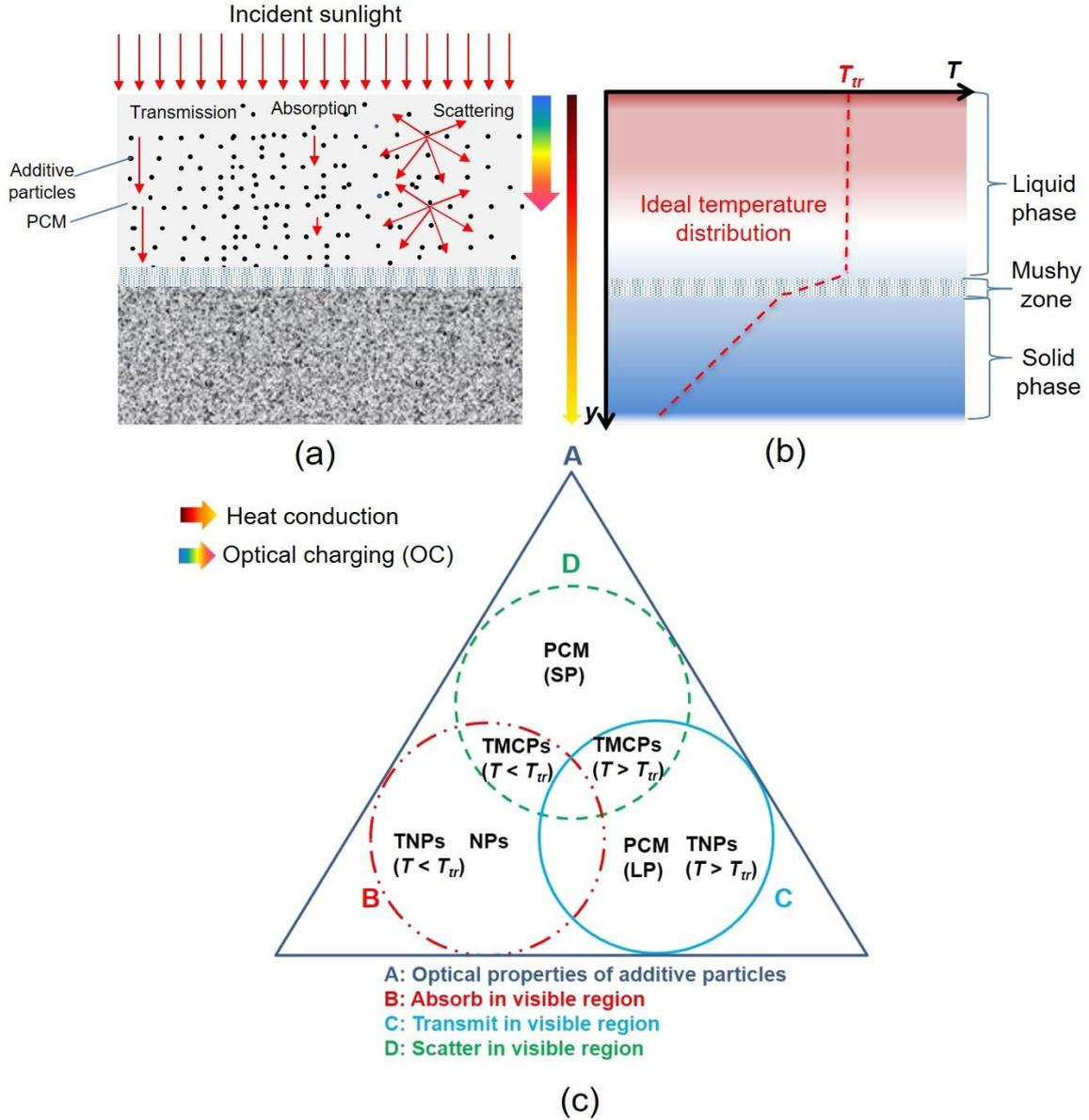

Fig. 2 Schematic diagram (a) showing the processes of photo-thermal energy conversion and melting during optical charging of the composite PCM, (b) representative ideal temperature field for accelerated thermostatic optical charging, and (c) optical properties of organic PCM (both in solid and liquid phases) and various additive particles.

## 3. MATERIALS AND METHODS

### 3.1 Synthesis routes

Organic PCM (OM46®) employed in the present work has been procured from Pluss® Advanced Technologies (India). In its pristine form, it is optically opaque and transparent in solid and liquid phases respectively. Essentially, two composite PCMs were synthesized, viz., non-thermochromic nanoparticles laden PCM (here on referred to as NP-PCM) and thermochromic microcapsule particles laden PCM (here on referred to as TMCP-PCM). For synthesis of NP-PCM, used-engine oil was first filtered; subsequently, desired amounts of it were mixed into molten heated pristine PCM (followed by ultrasonication) to form various concentrations (0.17 v/w%, 0.49 v/w%, and 0.83 v/w%) of NP-PCMs (see Fig. 3(a)). For synthesis of TMCP-PCM, thermochromic microcapsule particles (procured from Americos

Industries Inc.®, India. Chemical composition: Polyoxymethylene melamine, ODB -II, Methyl stearate, transition temperature, $T_{tr}$ = 55 °C,) of desired masses were mixed into molten heated pristine PCM (followed by ultrasonication) to form various concentrations (0.05 w/w%, 0.10 w/w%, 0.20 w/w%) of TMCP-PCMs (see Fig. 3(b)).

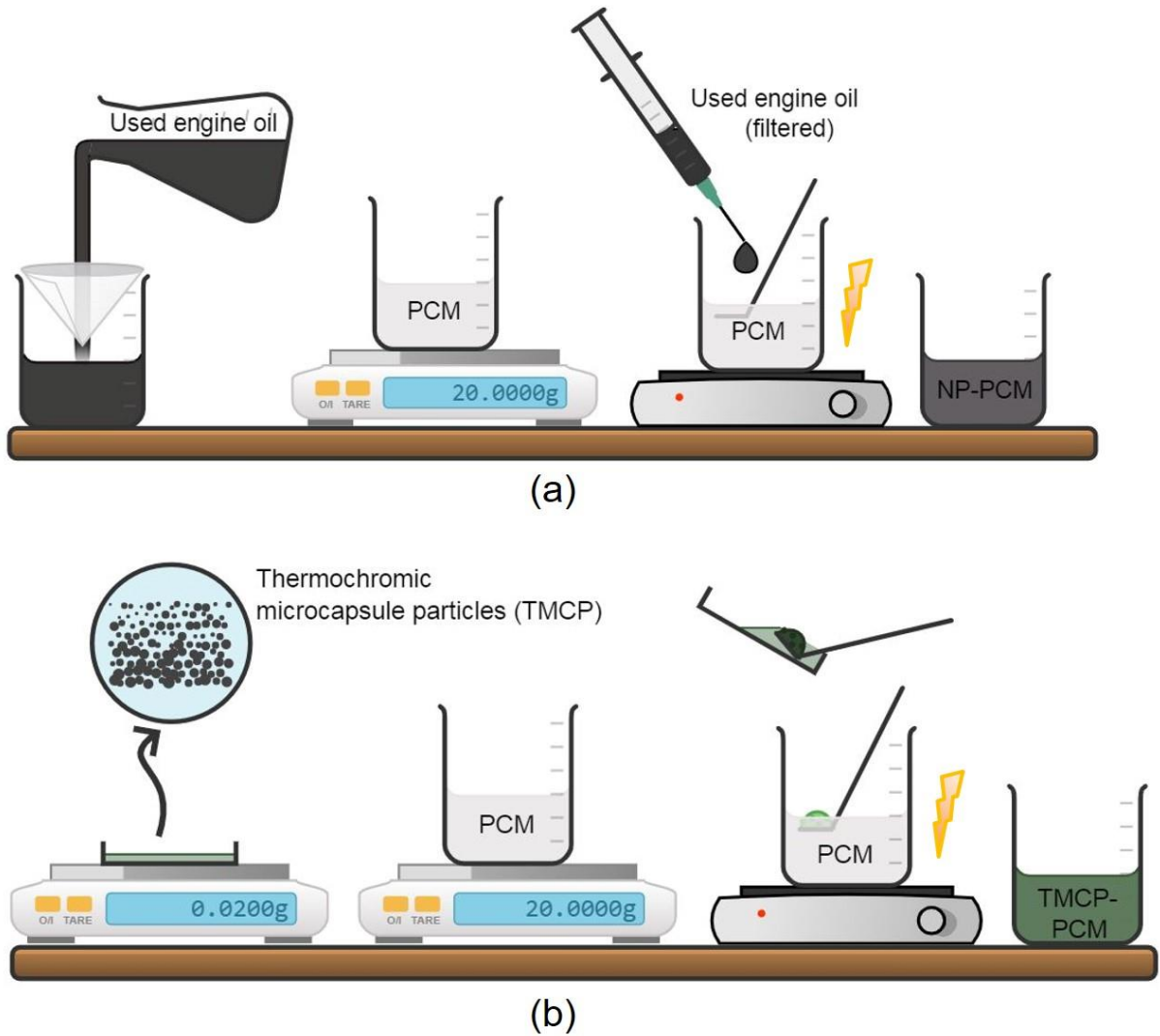

Fig. 3 Schematic diagrams showing the steps of synthesising (a) NP-PCM, and (b) TMCP-PCM

**3.2 Morphology, elemental composition and thermo-physical property characterization**
Scanning electron microscope (SEM) images of pristine PCM, NP-PCM TMCP-PCM, TMCP and transmission electron microscope (TEM) image of used engine oil nanoparticles (NP) are shown in Figs. 4 (a) – 4 (e). Clearly, carbon soot particles are amorphous and nanosized, whereas, thermochromic microcapsules are typically micron sized spherical particles (with diameter ranging between 1 μm to 10 μm).
Thermo-physical properties such as melting/freezing point, latent heat of fusion, and density values of pristine PCM (OM46, as provided by the manufacturer) are shown in Fig. 4 (h).

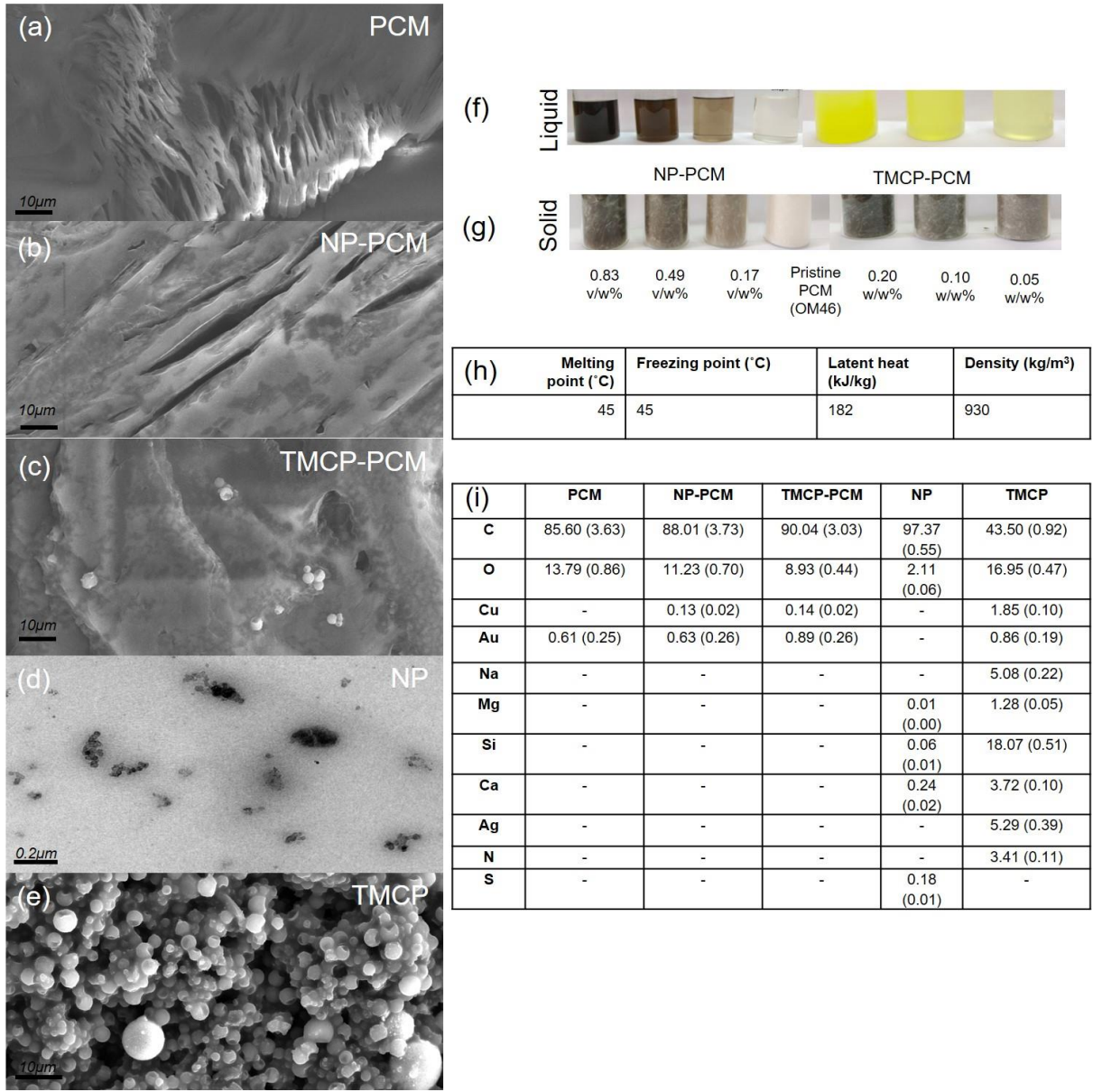

Fig. 4 Electron microscope images of (a) pristine PCM, (b) NP-PCM (0.83 v/w%), (c) TMCP-PCM (0.20 w/w%), (d) NP, and (e) TMCP; pictures of NP-PCM, pristine PCM, TCMP-PCM of different concentrations in (f) solid and (g) liquid phases; (h) thermo-physical properties of pristine PCM (OM46), and (i) composition of pristine PCM, NP-PCM, TCMP-PCM, NP, TCMP found through energy-dispersive X-ray spectroscopy (EDS).

Thermal conductivity measurements were made using transient plane source (TPS) method. Two identical cylindrical samples for each concentration of NP-PCM and TMCP-PCM were prepared. For measurement of thermal conductivity of a particular concentration of composite PCM, its two identical samples were kept on the either side of a hot plate which acted as a heater as well as a temperature sensor (TPS 500, Hot Disk®, measurement accuracy and measurement reproducibility as 5% and 2% respectively). Herein, through measurement of temperature change as a function of time lends determination of the thermal conductivity. Due to very low concentrations of additive particles employed in the present work, the thermal conductivity values of composite PCMs were similar to that for pristine PCM (see Fig. 5).

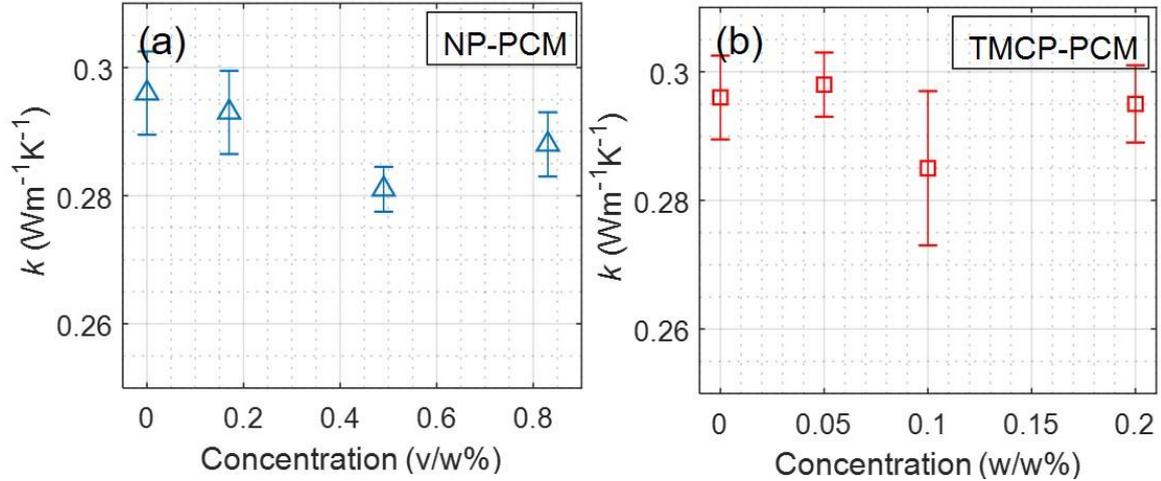

Fig. 5 Thermal conductivity for various concentrations of composite PCMs (a) NP-PCM, and (b) TMCP-PCM.

## 4. RESULTS AND DISCUSSIONS

### 4.1 Effect of phase change and thermochrism on optical properties of pristine and composite PCMs

Energy-dispersive X-ray spectroscopy (EDS) reveals that both the composite PCMs (i.e., NP-PCM as well as TMCP-PCM) have high carbon content (see Fig. 4 (i)). However, the particle size difference renders NP-PCM and TMCP-PCM to display markedly different optical properties. Whereas, NP-PCMs are weakly/strongly absorbing (depending on the concentration); on the other hand, TMCP-PCMs (after $T > T_{tr}$) essentially scatter weakly or strongly depending on the concentration (see Fig. 4(f) and 4 (g)).

Furthermore, with the objective of understanding the effects of phase and temperature on the optical properties of composite-PCMs (NP-PCM and TMCP-PCM), transmittance measurements were also made at different temperatures.

Essentially, a two-sided transparent cuvette was filled with the composite-PCM and kept over the hot plate. Once the temperature reached 70 °C, the heater was switched off. Light was made to strike on one of the transparent faces such that light after passing through the cuvette (composite PCM) falls on the spectrometer (Sekonic C-800 Spectro Master) detector (see Fig. 6 (d)). It may further be noted that the relative positions (distances) of the light source (LED), cuvette (containing the composite-PCM) and the detector (spectrometer) were so decided that spectrometer sensor is entirely covered by the light that is able to pass through the composite-PCM in the cuvette. Temperature measurements were made continuously throughout the cooling from 70°C to room temperature using an infrared imaging camera (Keysight U5850 True IR Thermal Imager). Also, snapshots were taken using a digital camera at different instants in time (see Fig. 6)

Figures 6(a), (b), and (c) show optical characteristics as a function of temperature for pristine PCM, NP-PCM, and TMCP-PCM respectively. In case of pristine PCM, in the liquid phase, it is highly transparent (as could be seen from the magnitude of the intensity measured by the spectrometer). However, as the PCM starts to cool down, crystallization starts and the magnitude of the intensity reaching the detector suddenly drops down and nearly reduces to zero after the entire PCM crystalizes (solidifies) completely. The aforementioned observation may be understood from the fact that as the crystallization progresses, the number and size of crystals (that scatter light) increase. This results in diffuse reflectance and transmittance of the incident light – hence only a negligibly small portion of the incident light is able to reach the detector (see Fig. 6 (a)).

In case of NP-PCM, in the liquid phase, due to presence of broad absorption carbon soot particles, most of the incident light gets absorbed within the PCM and rest which is able to escape through the NP-PCM and maintain its directionality is able to strike the detector. Similar to the pristine PCM, here also as the temperature drops, the crystallization starts and results in further reduction in magnitude of the intensity reaching the detector (see Fig. 6 (b)).

In addition to the absorption and crystallization effects, optical behaviour in case of TMCP-PCM is further complicated due to the presence of thermochromism. As the temperature decrease, the optical properties of both the base PCM and TMCPs start to change owing to crystallization and thermochromism respectively. Crystallization increases the magnitude of scattering and TMCPs also start to transform their characteristics from being highly absorbing to highly scattering. Hence, due to significant increase in the scattering albedo with decrease in temperature, the magnitude of the intensity reaching the detector significantly decrease and essentially reduces to negligible small value at room temperature (see Fig. 6 (c)).

Overall, for TMCP-PCM, the variations in optical signatures due to change in temperature are due to cumulative effects of phase change and thermochromism. On the other hand, in case of pristine PCM and NP-PCM, optical properties change with temperature predominantly due to phase change.

Furthermore, to clearly understand and quantify the roles played by thermochromism and phase change on optical properties, an additional set of optical characterization has been carried out. Herein, non-thermochromic nanoparticles (NP) and thermochromic microcapsule particles (TMCP) were dispersed in paraffin oil (PO) to form NP-PO and TMCP-PO dispersions respectively. It may be noted that, dispersions of similar concentrations (as employed in case of composite-PCMs) were prepared (see Fig. 7 (b)). As these are liquid dispersions, therefore, through careful optical characterization we could single out quantify the effect of thermochromism on optical properties. UV-Vis-NIR spectrophotometer (Cary 5000, Agilent Technologies) was employed to measure spectral direct transmittance ($T_{direct}$) both at room as well as elevated temperatures. It may further be noted that for elevated temperature measurements the liquid dispersion samples were heated to temperatures above the thermochromic transition temperatures. However, given the fact that the measurements across the entire UV-Vis-NIR wavelength band takes finite time, the temperature of the sample was continuously changing while the measurements were being recorded by the spectrophotometer. Figure 7 (c) clearly reveals that in case of NP-PO, optical properties are nearly invariant with respect to temperature. On the other hand, in case of TMCP-PO, the transmittance values are strong function of temperature (see Fig. 7 (d)). As mentioned earlier, for elevated temperature measurements, as the temperature was continuously changing and the measurement started from 1500 nm, therefor, most significant changes are observed in the NIR region. Relative lesser differences (between room and elevated temperature measurement values) in the visible region may be attributed to the fact that the due to time elapsed the sample's temperature has reduced to a value lower than the thermochromism transition temperature. In other words, the transmittance value starting from 1500 nm to 300 nm are actually at temperatures in between elevated and thermochromism transition temperatures.

The aforementioned analysis reaffirms that in case of pristine PCM and NP-PCM, the optical property change is solely due to phase change. Whereas, in case of TMCP-PCM, optical property changes are due to cumulative effects of phase change and thermochromism.

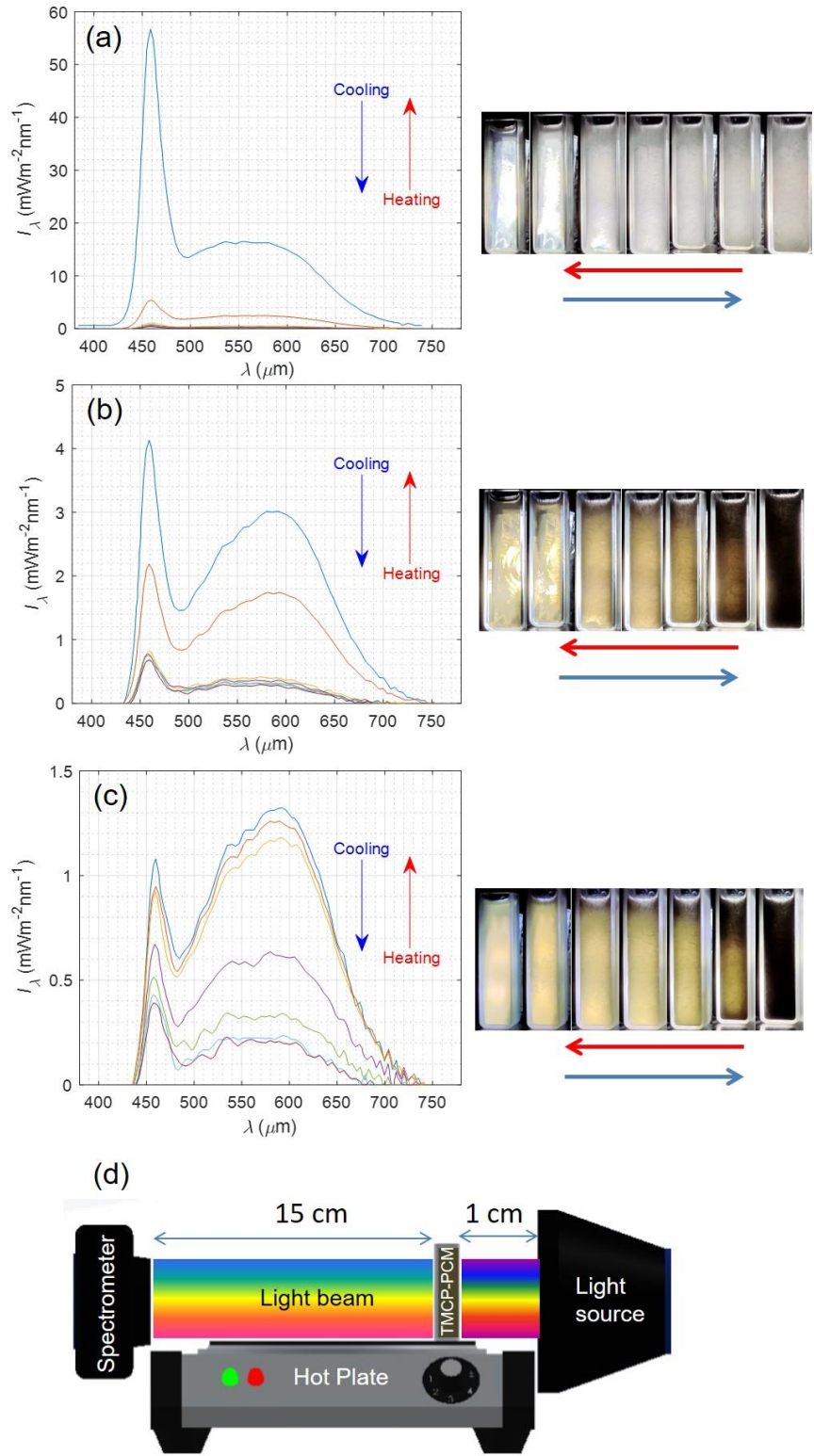

Fig. 6 Spectral intensity as a function of temperature for (a) pristine PCM, (b) NP-PCM, and (c) TCMP-PCM; (d) schematic diagram of the setup employed for taking optical measurements.

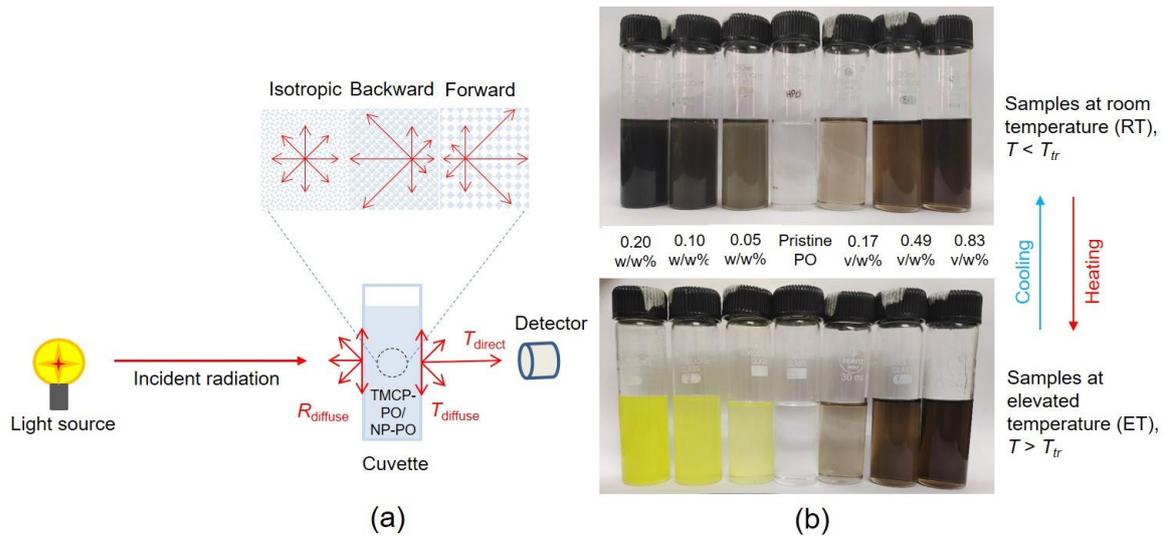
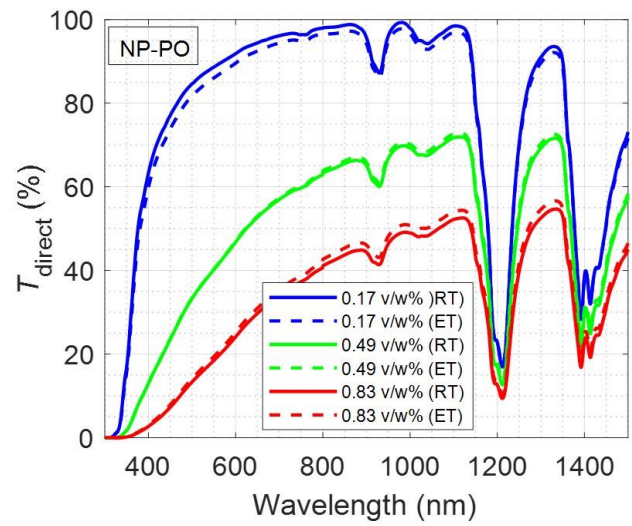
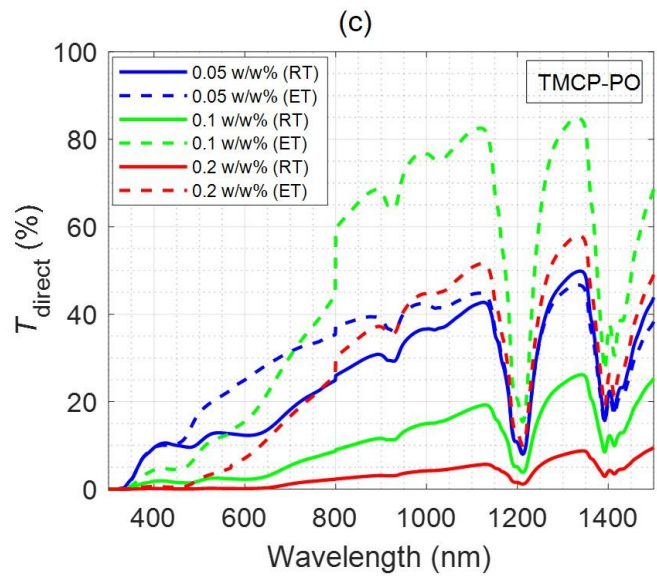

Fig. 7 (a) Schematic diagram showing the light-matter interaction and details of measurement configuration in the spectrophotometer, (b) photographs of pristine paraffin oil and various concentrations of NP-PO and TMCP-PO dispersions at room and elevated temperatures.

Spectral transmittance at room and elevated temperatures for (c) NP-PO dispersions, and (d) NP-TMCP dispersions.

## 4.2 Understanding photo-thermal energy conversion in composite-PCMs (NP-PCM and TMCP-PCM)

Now that we have understood and quantified the optical characteristics of pristine and composite-PCMs (NP-PCM and TMCP-PCM). Next, we investigate the effect of seeding thermochromic and non-thermochromic additives on photo-thermal energy conversion and hence melting rate of PCMs. For the said purpose, an experimental setup has been designed. Herein, a polypropylene cylindrical container was employed owing to the fact that polypropylene has low thermal mass and also holes can be drilled into it for inserting thermocouples along the depth direction.

To measure spatial temperature distribution, ten K-type thermocouples were inserted and fixed (using clear adhesive) along the height of the container (see Fig. 8 (a)). The container was filled with melted composite-PCM and allowed to solidify (and reach room temperature) in the container. Subsequently, the composite-PCM in the container was illuminated using a cold light source (halogen lamp (250 watts), connected to a light guide, spectra shown in Fig. 8(b)).

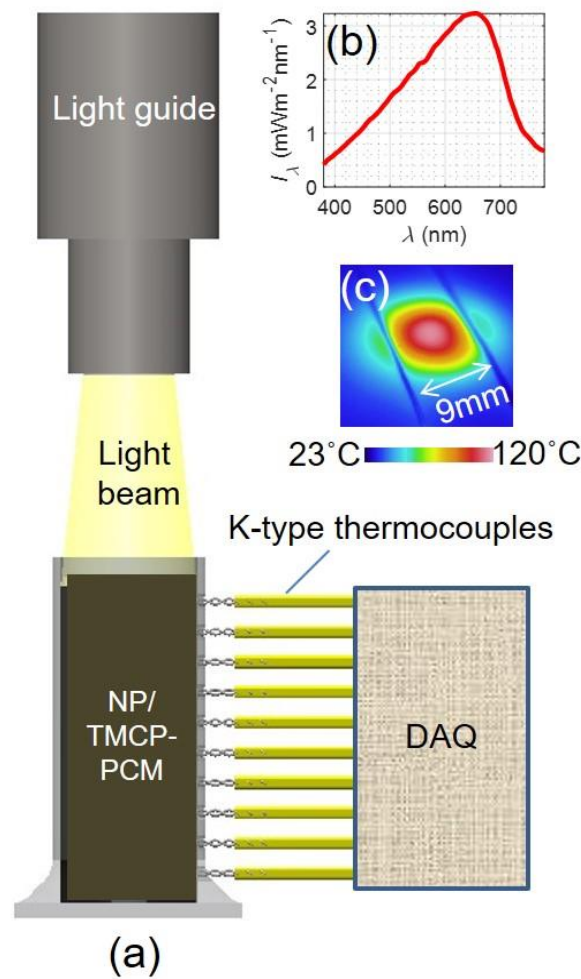

Fig. 8 (a) Schematic diagram showing the experimental setup employed for carrying out photo-thermal experiments, (b) spectra of the incident light, and (c) temperature distribution of the incident light on a plastic black surface.

Furthermore, spatial distribution of the light beam was also found by striking the light from the light guide onto a black plastic sheet (having low thermal conductivity). IR images were taken to get qualitative information regarding the uniformity of the light beam. Clearly, the temperature (and hence flux values) were not strong function of distance from the centre (up to approximately 8 mm) – hence nearly uniform illumination of the composite-PCM (see Fig. 8 (c)). Experiments were so designed that optical heating was done for first 20 minutes followed by 10 minutes of natural cooling under ambient conditions.

Spatial and temporal distribution curves (see Fig. 9 (a), (b), and (c)) clearly reveal that steady state is invariably reached for topmost thermocouples for all cases of TMCP-PCM. However, the subsequent thermocouples along the depth direction have yet not reached steady state. Furthermore, the highest temperature is nearly same irrespective of the concentration of TMCPs. On the other hand, in the case of NP-PCM (see Fig. 9 (d), (e), and (f)), steady state is not reached for any of the thermocouples. Furthermore, the peak temperatures are significantly impacted by the concentration of the carbon soot nanoparticles in the PCM. Peak temperatures increase with increase in nanoparticles concentration.

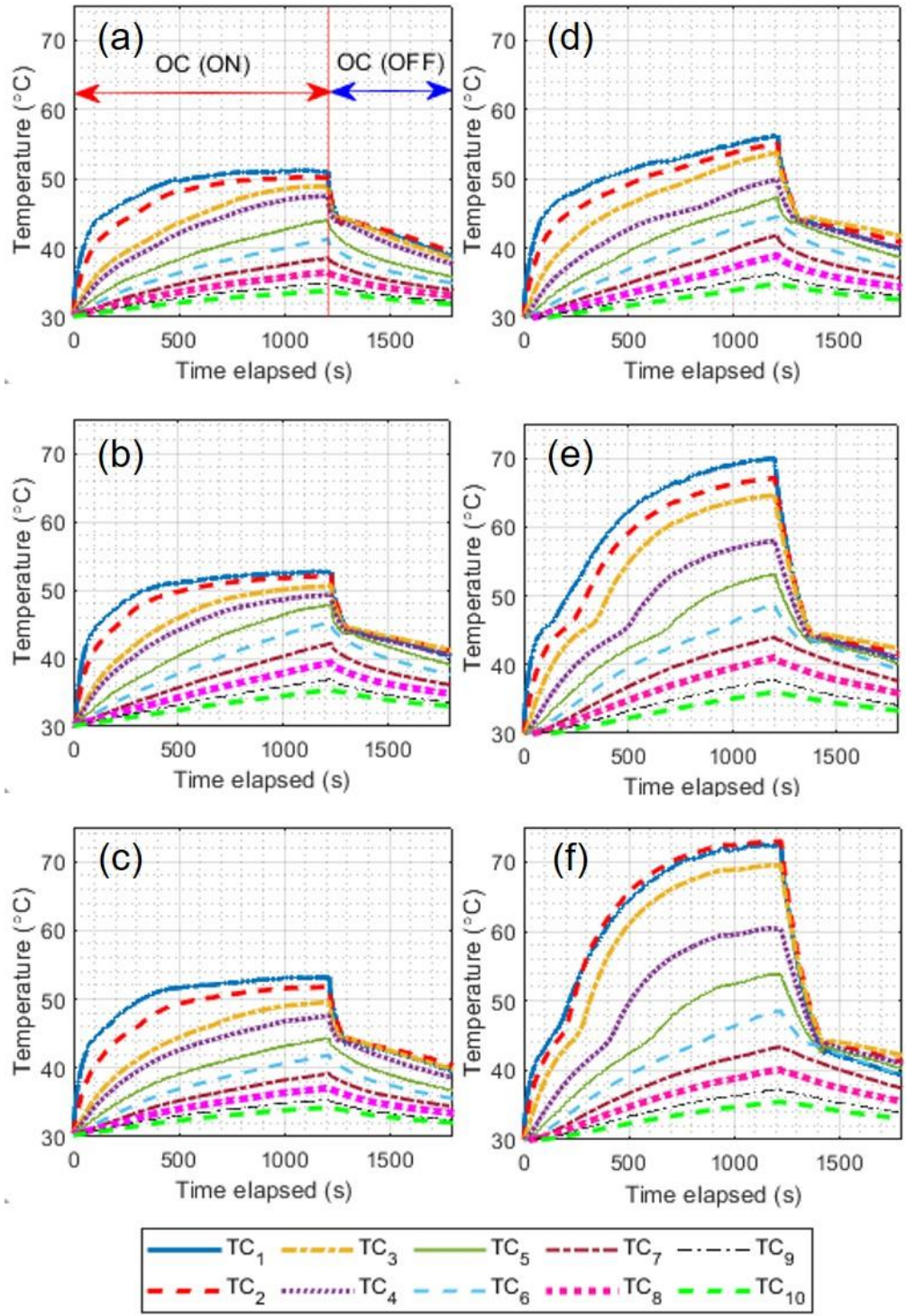

Fig. 9 Spatial-temporal temperature distribution for TMCP-PCM (a) 0.05(w/w%), (b) 0.10(w/w%) and (c) 0.15(w/w%); and NP-PCM (d) 0.17(v/w%), (e) 0.49(v/w%), and (f) 0.83(v/w%). Here, $TC_1$ through $TC_{10}$ are thermocouples along the depth direction, $TC_1$ is located at the top and $TC_{10}$ is located at the bottom.

Figure 10 details the spatial temperature distribution in various phases (liquid, mushy, solid) of composite-PCM just when the optical charging is turned off. Clearly, pronounced temperature spread is observed along the depth direction in case of NP-PCM (particularly in liquid phase) and peak values are strong functions of carbon soot nanoparticles concentration (see Fig. 10 (b)). On the other hand, in the liquid phase of TMCP-PCM, temperatures do not vary significantly in the depth direction. Also, the peak temperatures are almost constant and invariant with respect to TMCP particles concentration (see Fig. 10 (b)). In the solid phase, the temperature distribution is nearly linear (conduction being the predominant mode of heat transfer) in both TMCP-PCM and NP-PCM.

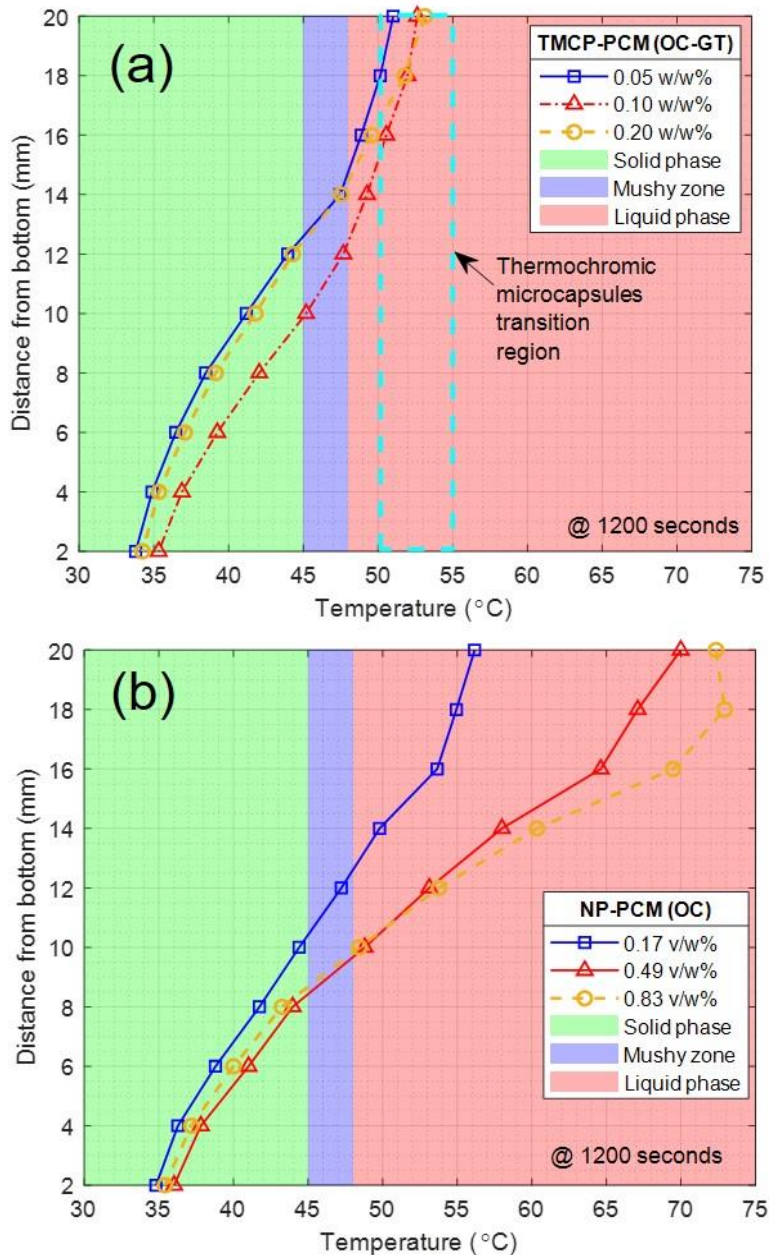

Fig. 10 Spatial temperature distribution (@ 1200 seconds) in various phases for various concentrations of (a) TMCP-PCM, and (b) NP-PCM.

### 4.3 Understanding melting front propagation in composite-PCMs (NP-PCM and TMCP-PCM)

Figure 11 shows melting front (defined @ 46 °C) propagation in case of NP-PCM. For a given concentration of carbon soot nanoparticles, the melting rate is highest in the initial stages of melting and the curve flattens out as the melting progresses. This may be understood from the fact that as the melting progresses, light has to transverse larger optical distance to reach the next solid layer. Furthermore, time required for the melting front to reach a particular value of *Y*, is directly related to the concentration of nanoparticles - least time is required for the NP-PCM with highest concentration of nanoparticles. This may be understood from the fact that photo-thermal energy conversion being confined to first few layers, highest peak temperatures are observed in NP-PCM with highest nanoparticles concentration. This further enhances the conduction heat transfer within the NP-PCM. At middle concentration of nanoparticles, the peak temperatures are not as high as observed in the case of NP-PCM with highest nanoparticles concentration. However, in this case, more volumetric photo-thermal energy conversion happens, which results in more accelerated melting in later stages of melting (see Fig. 11 (a), the melting front propagation curves for highest and middle concentrations tend to converge as the time elapses).

Comparing the melting front propagation curves for highest and lowest concentrations, we observe entirely opposite trend – the curves tend to diverge as the time elapses. The observed trends are essentially due to inefficient photo-thermal energy conversions in case of NP-PCM with lowest nanoparticles concentration – leading to low peak temperatures and hence reduced conduction heat transfer as well.

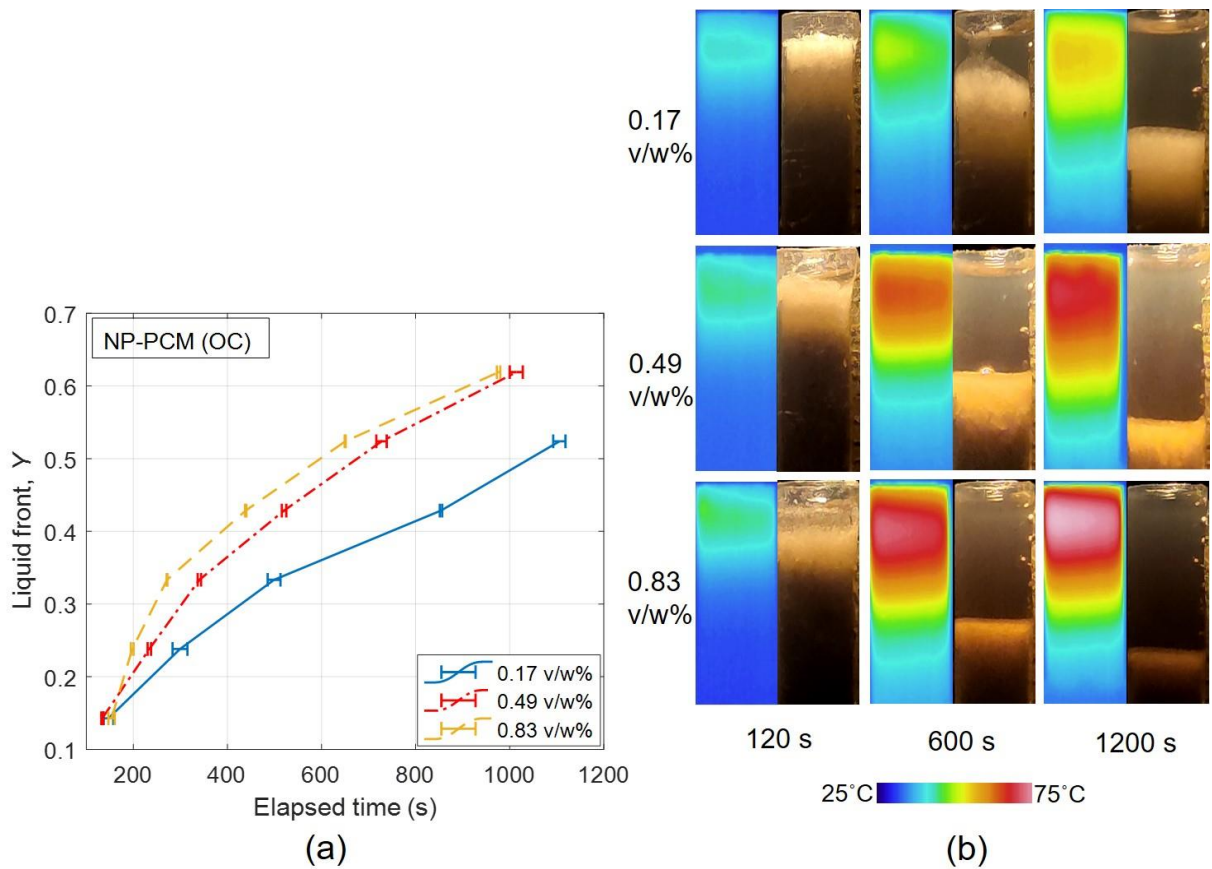

Fig. 11 (a) Melting front propagation as a function of time for various concentrations of NP-PCM, and (b) photographs showing melting front and various phases of NP-PCM during the melting process.

In case of TMCP-PCM, the melting front propagation curves reveal entirely different trends (see Fig. 12). Here, highest melting rates are observed for middle concentration (instead of highest concentration) of TMCP particles. In the initial stages of melting (i.e., *t* < 200 seconds), the highest concentration has the highest melting rate owing to better photo-thermal energy conversion in the solid phase. However, as the melting progresses, due to high concentration of TMCP particles in the liquid phase, and increase in scattering albedo at high temperatures (owing to thermochromism), the light is not able to reach the subsequent solid layers. Whereas, in the case of middle concentrations of TMCP particles, the number of TMCP particles are high enough to melt the solid PCM and low enough to allow the light to pass through and strike the subsequent solid layer of composite-PCM. Interestingly, for middle and lowest concentrations of TMCP particles, the melting rate accelerates in middle stages of melting owing to just the right mix of efficient photo-thermal energy conversion in the solid phase and acceptable optical path length in the liquid phase. Moreover, this lends the melting propagation curve of the lowest TMCP concentration to even overtake the melting propagation curve of the highest TMCP concentration in the final stages of melting.

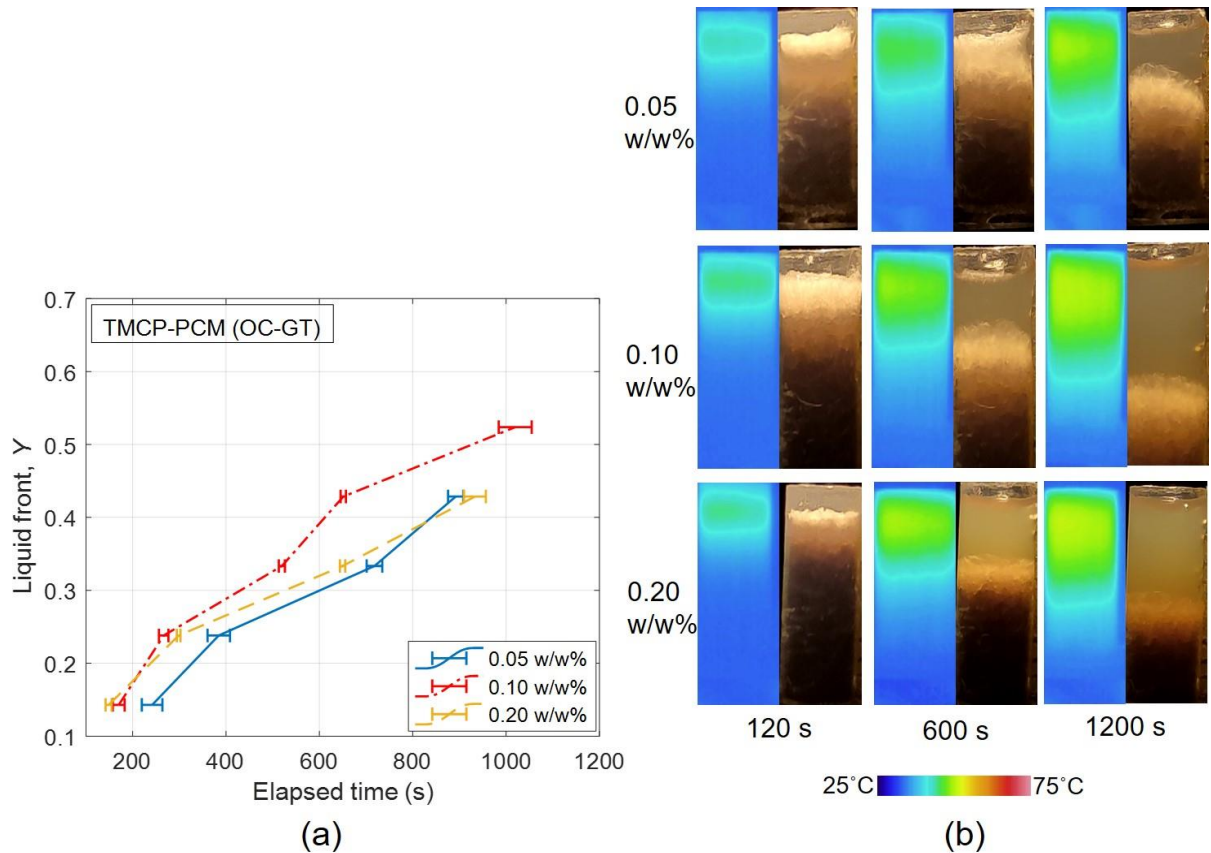

Fig. 12 (a) Melting front propagation as a function of time for various concentrations of TMCP-PCM, and (b) photographs showing melting front and various phases of TMCP-PCM during the melting process.

## 5. CONCLUSIONS AND FUTURE SCOPE

In the present work, optical charging of composite PCMs with and without thermochromism assistance has been experimentally analysed. Peak temperatures and optical pathlength in liquid phase are the key parameters that dictate the extent as well as rate of melting. Peak temperatures and optical pathlength in turn depend on optical properties, size and concentration of additive particles. In case of NP-PCM, the melting time is inversely related to particle concentration. Reduction in melting time on the order of 49 % is achieved as nanoparticles concentration is increased from 0.17 v/w% to 0.83 v/w%. On the other hand, in case of TMCP-PCM, melting time does not monotonically decrease with increase in particles' concentration. Instead, there exists an optimum concentration (0.10 w/w%) which allows for efficient photo-thermal energy conversion in the solid phase and at the same time permits passage through the liquid phase so that radiations strike the fresh solid layer. Herein, melting time reduction on the order 30 % is achieved relative to the lowest concentration (0.05 w/w%). Overall, opposed to optical charging of non-thermochromic-composite PCM; optical charging with thermochromism assistance allows for photo-thermal energy conversion at nearly thermostatic conditions. To further improve the melting rates, thermochromic nanoparticles (instead of TMCP particles) may be employed. This shall ensure lesser scattering and hence more optical clearance in the liquid phase – resulting in more accelerated thermostatic optical charging.

**ACKNOWLEDGEMENTS**
This work is supported by DST-SERB (under Sanction order no. CRG/2021/003272). Authors also wish to acknowledge the support provided by Department of Mechanical Engineering at Indian Institute of Technology Kanpur, Institute Instrumentation Centre (IIC) at Indian Institute of Technology Roorkee, Mechanical Engineering Department and School of Physics and Material Science at Thapar Institute of Engineering & Technology, Patiala.